# No Evidence for Particles
Casey Blood
Professor Emeritus of Physics
Rutgers University

## Abstract
There are a number of experiments and observations that appear to argue for the existence of particles, including the photoelectric and Compton effects, exposure of only one film grain by a spread-out photon wave function, and particle-like trajectories in bubble chambers. It can be shown, however, that all the particle-like phenomena can be explained by using properties of the wave functions/state vectors alone. Thus there is no evidence for particles. Wave-particle duality arises because the wave functions alone have both wave-like and particle-like properties. Further the results of the Bell-Aspect experiment and other experiments on entangled systems, which seem to imply peculiar properties for particles if they exist, are easily and naturally understood if reality consists of the state vectors alone. The linear equation-Hilbert space structure for the state vectors, by itself, can explain every mystery in quantum mechanics except the origin of the probability law.

## 1. Introduction
The centuries-old concept of particles is one of the cornerstones of our view of the structure of the physical universe. It has led to many insights and advances and is now so thoroughly accepted that it seems to be an indispensible feature of our conceptual landscape. In contrast to this apparent certainty, however, the *mathematics* of quantum mechanics, which gives an astonishingly accurate and wide-ranging quantitative description of nature, makes no mention of particles. Particles seem necessary, not to obtain the correct numerical answers—wavelengths, energies, cross-sections and so on—but rather to qualitatively account for observations that quantum mechanics, by itself, allegedly cannot explain.

Subjectively, it seems awkward to have a two-tiered scheme in which wave-function-based quantum mechanics determines all the numbers, while particles—absent from the quantum mathematics—supply the structure necessary for agreement with our perceptions. This suggests we take a close look to see if particles are really needed. And indeed we find that, in spite of all expectations, *particles are not necessary* to explain any observation. That is, there is no evidence that photons, electrons, protons and so on exist as particles, separate from the wave function—with "particle" being defined here as a carrier of mass, energy, momentum, spin and charge localized at or near a single point. All the particle-like properties can be explained by the properties of the wave function alone.

This is not the conclusion found in physics texts, however. If you look in a typical modern physics book, you will find analyses of experiments—particularly the photoelectric and Compton effects, and localized results from a spread-out wave

function—which reputedly prove particles are necessary for understanding the phenomena of physical existence. So why is there this contradiction between the physics texts and our conclusion here? It occurs because there are a number of particle-*like* properties of the wave function that are either not widely known, or their import is not appreciated.

There have been other attempts to show that one can describe all the particle-like phenomena of the physical world using wave functions alone [1–8], usually along the lines initiated by Everett [1] in his many-worlds interpretation. (See Appendix A for a comparison with Zeh's decoherence approach.) But these attempts do not explicitly show how all those phenomena are to be explained by the properties of the wave functions alone. So we will give here an explicit quantum mechanical explanation of all particle-like phenomena, thereby showing there is no evidence for particles.

We start in Sec. 2 by reviewing the primary evidence for particles, including the photoelectric and Compton effects, and localized effects from a spread-out wave function. In Sec. 3, we show, using group representation theory, that the particle-like properties of mass, energy, momentum, spin, and charge can be logically attributed to the wave function/state vector; it is not necessary to have particles that carry or possess these quantities. We then show in Sec. 4 that, even though quantum mechanics gives many simultaneously existing versions of reality, the theory implies that only one version will be (communicably) perceived. In Sec. 5 we show that a photon-like wave function spread out over many grains of film leads to the *perception* of the exposure of only one localized grain. We then show in Sec. 6 that, in contrast to the classical case, a small, localized portion of the wave function can transfer the full energy and momentum of the spread-out wave function to a localized electron (as in the photoelectric effect) or grain of film.

To illustrate the additional explanatory power accompanying the idea that there are no particles, we note in Sec. 7 that the presumed existence of particles often clouds the understanding of the outcome of experiments on entangled wave functions. We see, however, that the inferred peculiar properties of matter—instantaneous action at a distance in the Bell-Aspect case for example—become unnecessary if there are no particles.

Finally the conclusions are given in Sec. 8. It is seen that all the particle-like properties of matter are consequences of the properties of the wave functions. Wave-particle duality is just a division of the properties of the wave function into wave-like (primarily interference) and particle-like (mass, charge, spin, localization) properties. There is no need to have particles to explain the particle-like nature of the universe.

## 2. Alleged Evidence for Particles.

The photoelectric and Compton effects, both involving light-electron scattering, are exhibit A in attempting to show the existence of particles. Classically, light was considered to be a wave. But the photoelectric effect, in which electrons are ejected from a metal surface by shining light on it, could not be explained using classical ideas. The problem was that the electromagnetic wave was spread out over many billions of electrons in the metal, so each individual electron received (classically) only a very small amount of energy per second. In fact, using classical ideas, it should have taken days for an electron to gain enough energy to be ejected from the metal. Experimentally, the ejected electron current started almost immediately.

Einstein proposed that this could be understood if one assumed there was a localized particle, a photon, which was embedded within the light wave function and carried all the energy. This idea, using the Bohr formula $E = h\nu$ for the photon energy, was sufficient to account for all the data.

The photon idea was seemingly confirmed in Compton scattering. If one assumes (1) there is a particulate electron with relativistic energy $E = \sqrt{m^2 c^4 + p_e^2 c^2}$, (2) there is a particulate photon that carries energy and momentum $E = h\nu, p = h/\lambda$, and (3) that energy and momentum are conserved in a collision, then the correct equations describing Compton scattering can be derived.

Does this prove there are particulate photons and electrons? No it does not, because one can also derive the photoelectric and Compton effect formulas using only properties of the wave function, with no assumption that particles exist. There are two parts to the derivation. The first is to show from group representation theory that mass, energy, momentum, spin and charge can be logically attributed to the state vector. And the second is to show that, in contrast to the classical properties of waves, a small part of a spread-out wave function can transfer the full complement of energy and momentum to another, localized wave function.

Exhibit B in the alleged evidence for particles pertains to the localized effects of spread-out wave functions. For example, if a photon-like wave function goes through a single slit, becomes spread out and hits a screen covered with grains of film, one will find only a single grain of film exposed. Or if an electron-like wave function is scattered off a proton (wave function) so there is an outward spherical wave, and the detector is a sphere covered with film grains, again, only a single grain of film will be exposed. It is tempting to interpret these results as implying that there was a localized particulate photon or electron, embedded in the wave function, which hit and exposed just one grain. However, we will show that quantum mechanics, by itself, leads to the *perception* of only one localized grain exposed. And so the perception of localized effects from a spread-out wave function also does not provide evidence for the existence of particles. One can extend this argument to show that quantum mechanics, by itself, also predicts that we will perceive particle-like trajectories.

A third argument for the existence of particles arises because the wave function of quantum mechanics often contains several versions of reality, but we perceive only one. In the particle view, a particle rides along on just one version of the wave function, and it is the particled version that we perceive. We will show in Sec. 4, however, that quantum mechanics alone implies only one version is perceived, so perception of only one version of reality does not provide evidence for particles.

There are also chemical and thermodynamic arguments for the existence of particles. But these primarily use the idea of discrete units of matter, and the electron-like, nucleus-like, atomic-like, and molecule-like wave functions provide this discreteness just as readily as actual particles. So we will not pursue this line of evidence further.

## 3. The particle-like properties of mass, energy, momentum, spin, and charge.

In classical physics, it is assumed that particles possess the properties of mass, energy, momentum, spin (angular momentum), and charge. So if we wish to show there is no evidence for particles, we must show that these particle-like properties can be logically attributed instead to the wave functions/state vectors.

This is done using group representation theory. The equations of quantum mechanics are linear equations for the state vectors, with the equations being invariant under inhomogeneous Lorentz transformations (four-dimensional "rotations" plus translations) and internal symmetry group operations. Invariance under inhomogeneous Lorentz transformations implies that the solutions—the state vectors—can be labeled by mass, energy, momentum, and angular momentum and its z component [9]. In addition, invariance under internal symmetry groups [10-11] implies charges are also properties of the state vectors. So we see that the equations of quantum mechanics, using only the very general principles of linearity and invariance, imply that mass, energy, momentum, spin, and charge can be logically attributed to the state vectors.

Further, the linear operators corresponding to energy, momentum, z component of angular momentum, and charge are generators of the Lorentz or internal symmetry groups. This implies that in a direct product state, the values add algebraically. For example,

$$(P)_{op}\{|\ldots, p_1, \ldots\rangle |\ldots p_2, \ldots\rangle\} = (p_1 + p_2)|\ldots, p_1, \ldots\rangle |\ldots p_2, \ldots\rangle \qquad (1)$$

In addition, invariance under the symmetry group implies that the total energy, momentum, z component of spin, and charge are conserved in interactions.

In summary, linearity, invariance and group representation theory imply it is reasonable to assign the particle-like properties of mass, energy, momentum, spin and charge to the state vectors, with the correct addition and conservation laws then automatically holding. Thus neither the particle-like properties of mass, charge and so on, nor the addition laws, nor the conservation laws can be used as evidence for the existence of particles because these properties can be logically attributed to the state vectors. (Note that we are not trying to prove particles don't exist. We are only showing that the properties of state vectors can account for all the particle-like properties of matter.)

## 4. Quantum mechanics implies perception of a single version of reality.

The wave function often contains many simultaneously existing versions of reality, so we might expect quantum mechanics to predict the simultaneous perception of more than one version of reality. However, we can show (1) that quantum mechanics allows the perception of only a single version of reality, and (2) that classical consistency, exemplified primarily by if-then logic, is obeyed in the perceived state.

To illustrate, we use a Stern-Gerlach experiment on a spin 1 atom. After going through the magnet, the wave function of "the atom" consists of three different, non-overlapping parts that follow three different paths. There is a detector, D(i), on each of the three paths which can read "n" (no detection) or "y" (yes, detection). And there is an observer who perceives the readings on the three detectors. After the wave function passes through the magnet but before it reaches the detectors, the state vector is (with $|a(1)|^2+|a(2)|^2+|a(3)|^2 =1$)

$$|\Psi(1)\rangle = a(1)|1\rangle|D(1,n)\rangle|D(2,n)\rangle|D(3,n)\rangle|Obs, n, n, n\rangle \\ + a(2)|2\rangle|D(1,n)\rangle|D(2,n)\rangle|D(3,n)\rangle|Obs, n, n, n\rangle \\ + a(3)|3\rangle|D(1,n)\rangle|D(2,n)\rangle|D(3,n)\rangle|Obs, n, n, n\rangle \quad (2)$$

The objective is to show that quantum mechanics alone does not allow the perception of anything besides one classical reading—either (y,n,n) or (n,y,n) or (n,n,y)—of the detectors. To show this holds, we assume a certain characteristic for our sensory perceptions; we assume they are *communicable*. This weak assumption is certainly in agreement with our understanding of perceptions. Thus we have

> **The communicability criterion.** *If no version of the observer can communicably verify that anything other than a classical version of reality is perceived, then that is sufficient to declare that quantum mechanics allows only classical perceptions, in accord with our everyday experience.*

Note: Aside from probability, the tenets of the historically important "measurement theory" are being replaced here by basic principles of quantum mechanics—linearity, properties of the Hamiltonians, orthogonality arguments—and the criterion of communicability.

To implement this communicability criterion, we now ask the observer to write "1" on a piece of paper if she sees only y,n,n; "2" if she sees only n,y,n; "3" if she sees only n,n,y; and "4" if she sees any other, non-classical result. Then after the wave function hits the detectors and the observer perceives and writes down the readings, the state vector is

$$|\Psi(2)\rangle = a(1)|1\rangle|D(1,y)\rangle|D(2,n)\rangle|D(3,n)\rangle|Obs\ sees\ y,n,n\rangle|paper\ 1\rangle \\ + a(2)|2\rangle|D(1,n)\rangle|D(2,y)\rangle|D(3,n)\rangle|Obs\ sees\ n,y,n\rangle|paper\ 2\rangle \\ + a(3)|3\rangle|D(1,n)\rangle|D(2,n)\rangle|D(3,y)\rangle|Obs\ sees\ n,n,y\rangle|paper\ 3\rangle \quad (3)$$

There are several remarks concerning this equation.
**(1)** The three branches of the wave function are orthogonal and remain so for all time. This implies each of the three branches evolves as if the others were not there. There can be no information passed between branches so that each version of the observer in Eq. (3) can only be aware of events on its own branch. See Appendix B, on the no-interaction rule, for a justification of these remarks.
**(2)** The particle-detector, detector-observer, and observer-paper interaction Hamiltonians guarantee there will be the classically expected "if-then" agreement of the atomic, detector, observer, and paper states on each branch: if the version of the atomic state is *j*, the version of the detectors will indicate state *j*; if the version of the detectors indicates state *j*, the version of the observer will perceive state *j*; and if the version of the observer perceives state *j*, the writing on the paper will indicate state *j*.  Also because a classically inconsistent reading on the detectors never occurs within a single branch, a version of the observer never perceives a non-classical result such as D(1,y)D(2,y)D(3,n).
**(3)** Although we do not explicitly show it here, it should be obvious that the interaction Hamiltonians plus the no-interaction rule guarantee that the perceptions of the versions of multiple observers always agree within a branch.
**(4)** As indicated in (2), the interaction Hamiltonians force the form of Eq. (3).  No special basis was chosen in writing that equation.
**(5)** The non-classical designator "4" is never written in Eq. (3) so that no matter what basis is chosen for the observer states, only classical results will be perceived.

### Perception of a single version of reality.

Because it is important, we will expand on the last point by examining a case where one might *expect* a non-classical perception.  Suppose after the observer perceives but before she writes anything down, we consider the quantum mechanically allowed observer state

$$\sqrt{2}|\Psi,+\rangle = |\text{Obs sees y, n, n}\rangle + |\text{Obs sees n, y, n}\rangle \qquad (4)$$

On the surface, this state seems to imply the perception of both y,n,n and n,y,n.  But that inference is not correct; it does not lead to the simultaneous perception of both options if we use the "communicable" understanding of perception.

To see this, note first that the two components of this state are in different, non-communicating universes because the set of firing neurons is different in the two cases, and the quantum state of a firing neuron is different from, and orthogonal to, that of a non-firing neuron. (The ion densities inside and outside the neuron are different for the firing and not-firing states, so the two types of states must be orthogonal.)  Thus each component of the version of the observer in Eq. (4) will evolve in time entirely independently of the other component.  We now have the observer write down what she perceives and include the written-on paper in the state.  Then the dynamics of the observer's brain-body, plus the non-interaction of the two versions imply that the y,n,n version of the observer must write "1" and the n,y,n version must write "2;"

$$\sqrt{2}U(t)|\Psi,+\rangle$$
$$= |\text{Obs sees y, n, n}\rangle|\text{paper 1}\rangle + |\text{Obs sees n, y, n}\rangle|\text{paper 2}\rangle \qquad (5)$$

"4", implying "I perceive *both* y,n,n *and* n,y,n" is never written.  Generalizing from Eqs. (4) and (5), we see that *no* basis vector leads to the communicable perception of a non-classical result (a 4); only the classical 1, 2, or 3 will be written.  There is no overall observer, residing in both the y,n,n and the n,y,n universes, that (communicably) perceives both parts. (Note: One can somewhat trivially avoid the communicability assumption by putting our conclusion in the form: "If a version of the observer perceives a non-classical result, then she cannot communicate it; only classical, single-version states of reality are communcably perceived.")

To put this another way, there is a superselection rule that applies to the states corresponding to the different branches. Such a rule holds when there is no interaction between different states for all time, as will be true for the branches of the wave function for virtually all types of detector-recorders. The superselection rule essentially says there is no point in considering linear combinations of states that never interact because none of the (communicably) perceived results are different from when the states are considered separately.

Zurek [2,3, 24] also arrives at the conclusion that only one classical result is perceived—either y,n,n or n,y,n or n,n,y—but he uses a method which seems unduly complicated. His use of Quantum Darwinism, Shannon entropy, and multiple records in the environment seems necessary only if one ignores the fact that the state of Eq. (4) does *not* correspond in any meaningful way to a version of the observer perceiving both y,n,n and n,y,n. Since there is no reason to ignore this fact, Zurek's method can be bypassed, at least for the purpose of showing that only classical results are perceived in quantum mechanics.

Pure quantum mechanics and probability.
The quantum mechanics we are dealing with here might be called pure quantum mechanics— defined as no particles, no hidden variables, no collapse, and no *a priori* probability. Only the wave function exists, with all its multiple versions of reality. This is Everett's [1] starting point for his many-worlds interpretation. The conclusion we draw from Secs. 3-7 is that pure quantum mechanics does an excellent job of describing our perceptions.

But pure quantum mechanics seems to have a flaw; it is apparently not capable of accounting for the probability law. Why do we suspect this? Because every version of the observer perceives its respective outcome with 100% probability on every run, *regardless of the values of the coefficients,* so it would seem there can be no *coefficient-dependent* probability of perception, in contradiction to experiment. A full discussion of this point is, however, outside the scope of this paper.

## 5. The Perception of Localization.

A major property of particles in the classical view is that they produce localized effects. However, we will show here that the mathematics of quantum mechanics alone implies spread-out wave functions produce the *perception* of localized effects. So localization cannot be considered as evidence for particles (or hidden variables, or collapse).

We consider the case of a spread-out light wave function hitting a screen covered with N grains of film. The wave function-film grain Hamiltonian will be the sum of N terms,

$$H = \sum_{i=1}^{N} h(i) \tag{6}$$

where $h(i)$ is the interaction between the wave function and the $i$th grain. Eq. (6) implies that after the light wave hits the grains and is annihilated, the state of the grains and observer is also a sum of N terms

$$|\Psi\rangle = \sum_{i=1}^{N} a(i)|\text{gr } 1\rangle \ldots |\text{gr } i\rangle^* \cdots |\text{gr } N\rangle |\text{Obs sees grain } i \text{ exposed}\rangle \qquad (7)$$

where the $a(i)$ has to do with the amplitude of the light wave at grain $i$, and the asterisk denotes an exposed grain. (The small part of the wave function that hits grain $i$ can transfer the full energy of the wave to the grain and expose it; see Sec. 6). That is, the sum form of the Hamiltonian in Eq. (6) implies that the final state will be a sum of branches in which one and only one grain is exposed on each branch of the state vector.

Further, the arguments of Sec. 4, plus the orthogonality of the unexposed and exposed states of the grains, guarantee (1) that there will be N branches, (2) that there will be a version of the observer on each branch, (3) that that version will perceive only what occurs on that branch—which is the exposure of one and only one grain, and (4) that no linear combination of the versions will perceive more than one exposed grain. Thus, even though the wave function is spread out over many grains, each version of the observer will (communicably) perceive precisely one grain exposed. So we do not need particles (or hidden variables, or collapse) to justify the *perception* of one and only one localized grain being exposed by a spread-out wave function.

Suppose next we consider the case of an electron wave function scattered off a proton wave function, with the detector being a sphere coated with grains of film. Then exactly the same argument applies: the Hamiltonian will be of the sum form of Eq. (6); the final wave function will therefore be a sum of terms with one and only one grain exposed in each term; and no version of the observer will (communicably) perceive more than one grain exposed. Thus even though it certainly appears to each version of the observer that a particulate electron bounced off a particulate proton and hit one grain, that picture is not needed to explain the perception of one and only one exposed, localized grain.

What about particle-like trajectories? An extension of the above argument, with many layers of film grains and with the particles going through the film grains, shows that quantum mechanics always predicts the perception of one and only one smooth trajectory. A disjoint trajectory does not occur on any single branch of the wave function and so it will never be perceived (because perception is only within a single branch). This is in line with the classical if-then structure discussed in Sec. 4.

An extension of the localization reasoning also shows that even if a macroscopic object has a center of mass wave function spread out over, say, a meter, it will be *perceived* as being highly localized.

Thus the particle-like property of localization can be entirely accounted for by the perception of one and only one version of reality in quantum mechanics, so localized effects cannot be used as evidence for particles.

# 6. Energy Transfer.
# The Photoelectric and Compton Effects.

We now consider the problem that prompted Einstein to propose the existence of particles of light, namely the transfer of energy from a spread-out light wave to a single localized electron (wave function). In the classical picture of the photoelectric effect, because a spread-out light wave hits many electrons, and because a small part of the light wave is presumed to carry only a small part of the energy, only a very small part of the energy of the wave can be transferred to the electron [12]. We will show that is not true in quantum mechanics, however; a "small part" of the incoming photon-like wave function can transfer the full energy of the wave function to a localized particle (particle-like wave function).

We will initially use film grains instead of electrons. To start, suppose we have a single grain and we shoot a photon-like wave function at the grain, focused so the entire wave function hits the grain. We assume the wavelength of the photon is short enough to expose the grain, so the time evolution, denoted by U(t), will be

$$|\text{ph}, \lambda\rangle |\text{grain}\rangle \rightarrow U(t)[|\text{ph}, \lambda\rangle |\text{grain}\rangle] = |\text{grain}\rangle^* \tag{8}$$

with the exposed grain, indicated by the asterisk, absorbing the full energy of the localized photon-like wave function. Now suppose the single photon-like wave function is divided into two parts that take separate trajectories,

$$|\text{Ph}\rangle = a(1)|\text{ph}(1), \lambda\rangle + a(2)|\text{ph}(2), \lambda\rangle, \tag{9}$$
$$|a(1)|^2 \ll 1, |a(1)|^2 + |a(2)|^2 = 1$$

with the "weak" part 1 hitting grain 1 and the "strong" part 2 hitting grain 2. Then the final state is

$$\begin{aligned} U(t)|\text{Ph}\rangle &= U(t)\{[|a(1)|\text{ph}(1), \lambda\rangle + a(2)|\text{ph}(2), \lambda\rangle] \, [|\text{gr 1}\rangle |\text{gr 2}\rangle]\} \\ &= a(1)U(t)\{|\text{ph 1}\rangle |\text{gr 1}\rangle |\text{gr 2}\rangle\} + \\ &\quad a(2)U(t)\{|\text{ph 2}\rangle |\text{gr 1}\rangle |\text{gr 2}\rangle\} \\ &= a(1)|\text{gr 1}\rangle^* |\text{gr 2}\rangle + a(2)|\text{gr 1}\rangle |\text{gr 2}\rangle^* \end{aligned} \tag{10}$$

Lines 2 and 3 follow from the linearity of U(t) while line 4 follows from Eq. (8), applied twice, and the fact that the "1" part of the photon wave function can expose only grain 1 and the "2" part only grain 2.

This argument can obviously be extended to a wave function spread out over N grains, with the small fraction of the wave that hits the *i*th grain exposing that grain, regardless of the magnitude of the coefficient *a*(*i*). The "amount," $|a(i)|^2$, of the light wave function hitting grain *i* only has to do (in quantum mechanics) with the *probability* of that grain being exposed. There is no *gradual* accumulation of energy by the grains as there is in the classical picture.

In the photoelectric effect, the analogous process is the photon-like wave function hitting many electrons. But the mathematics is just the same; the photon-electron Hamiltonian is a sum of parts, one for each electron, and the time translation operator U(t) is linear. Thus the end result will be a sum of terms in each of which a *single electron* (electron-like wave function) has absorbed an amount of energy that can be as large as the full energy ($h\lambda$) of the photon wave function. There is no sharing of the energy by a large number of electrons. This leads to agreement with the experimental results in the photoelectric effect, even though the existence of particulate photons or electrons was not assumed.

In the Compton effect, we assume, in accord with Sec. 3, that the energy and momentum belong to the photon-like (m=Q=0, S=1) and electron-like (m=m$_e$, Q=$-e$, S=1/2) wave functions, and that total energy and momentum are conserved. This, plus the above energy-transfer argument (which also applies to momentum) are sufficient to derive the experimentally verified Compton effect equations.

## 7. The Bell-Aspect experiment.

There are a number of experiments involving entangled wave functions—for example the Bell-Aspect experiment [13,14], Wheeler's delayed-choice experiment [15,16], interaction-free measurements [17,18], and the quantum eraser [19,20]—where the results are difficult to understand if one assumes the physical world is made of particles, but they follow simply from no-particle quantum mechanics. To illustrate, we will analyze, the Bell-Aspect experiment, which is a descendant of the arguments of Einstein, Podolsky, and Rosen [21].

Two photons are (nearly) simultaneously emitted from an atom which both before and after the emission is in a spin 0 state. One photon travels to the left and the other to the right, so the zero-spin state of the photons is (assuming the same set of coordinates are used for the two photons)

$$\sqrt{2}|\Psi\rangle = |+, \text{left}\rangle|-, \text{right}\rangle - |-, \text{left}\rangle|+, \text{right}\rangle \qquad (11)$$

If the right beam of photons is split into a + polarization and a – polarization by an apparatus oriented at angle 0 and the left beam is split into a + polarization and a – polarization by an apparatus oriented at an angle $\theta$ then the wave function can be re-expressed as

$$\sqrt{2}|\Psi\rangle = [\cos\theta|+, \text{left}\rangle + \sin\theta|-, \text{left}\rangle]|-, \text{right}\rangle \qquad (12)$$
$$- [-\sin\theta|+, \text{left}\rangle + \cos\theta|-, \text{left}\rangle]|+, \text{right}\rangle.$$

For each pair of photon wave functions that go through the apparatus, there are four possible outcomes. These, along with the probabilities predicted by quantum mechanics, are

$$\begin{array}{ll} +, L: +R & \sin^2\theta/2 \\ +, L: -R & \cos^2\theta/2 \\ -, L: +R & \cos^2\theta/2 \\ -, L: -R & \sin^2\theta/2 \end{array} \qquad (13)$$

11The probabilities are experimentally confirmed in the Aspect experiment.

But now suppose we consider the experiment from the classical particle point of view. Then there would be one photon moving to the left in a definite polarization state and another moving to the right, also in a definite polarization state. It was Bell's stroke of genius to find a way to show that the four probabilities in Eq. (13) could not hold in the classical picture in which each of the two localized photon particles possesses a definite state of polarization which is not changed by a measurement on the other, distant photon.

Suppose, however, that one insists on the classical picture of localized carriers of the particle-like properties. Then to account for Eq. (13) holding experimentally, one must postulate that there is some unknown force which instantaneously changes the polarization state of the second particle when the first is measured [22]. But it is not necessary to postulate such a peculiar force if one assumes only the state vectors exist; quantum mechanics perfectly predicts the results (Eq. (13)) without the presumption of an instantaneous-action-at-a-distance force. The long-range correlations between the two photon states implied by Eq. (13) are built into the entangled wave function.

## 8. Summary and Conclusions.

The concept of particles, although widely accepted, does not occur in the mathematics of quantum mechanics, where the equations are only for the wave functions. So it seems prudent to review the evidence for particles. There are four primary reasons to believe they exist. First, quantum mechanics predicts that in many circumstances there will be several versions of reality. But we perceive only one, with the usual presumption being that the perceived version is made up of particles. Second, it is normally presumed that localized particles are carriers of the particle-like properties of mass, energy, momentum, spin, and charge. Third, a spread-out wave function produces a localized effect, presumably caused by a localized particle embedded in the wave function. And finally, the results of the photoelectric and Compton effects are nicely explained by assuming there are particulate photons and electrons which carry the full energy and momentum of the wave function.

We find, however, that all four of these observations can be explained *within quantum mechanics*, without assuming the existence of particles, by a careful examination of the properties of the wave function and the way in which it is perceived. These properties are derived from just a few basic principles: (1) linearity; (2) invariance under the inhomogeneous Lorentz group and the internal symmetry group; (3) the existence of a scalar product for kets; and (4) minimal assumptions about the interaction Hamiltonians, primarily that a wave function on path *i* can only trigger a detector on path *i* and the observer perceives "yes" if the detector reads "yes". These four, plus minimal assumptions about orthogonality, lead to (5) the orthogonality of, and non-communication between different branches of the wave function. Finally, to show the quantum mechanical mathematics leads to our "classical," single-version perceptions, it is sufficient to assume (6) that the results of perception are *communicable*.

Using these wave-function-only principles, one can show first that more than one version of the wave function is never communicably perceived, so that perception of a single version of reality does not offer evidence for particles (or hidden variables, or collapse). Second, one can show from group representation theory that mass, energy, momentum, spin and charge can be logically attributed to the state vector, so these particle-like properties also provide no evidence for particles. Third, quantum mechanics, by itself, predicts that a spread-out wave function will be *perceived* as triggering one and only one localized detector—a grain of film, for example. And finally, the photoelectric and Compton effects can be explained using properties of the wave function alone, without resorting to the concept of particles.

All the other evidence for particles can be explained in terms of properties of the wave function using essentially the same reasoning as was used in these four, so *all the particle-like properties of physical existence follow from the equations and principles of quantum mechanics alone*. Thus, since quantum mechanics has co-opted all the reputed evidence for particles, we find there is no evidence for their existence. It is not necessary to assume matter is constructed from some peculiar amalgam of particles and waves. Instead, wave-particle duality is simply a duality in the properties, wave-like *and* particle-like, of the wave function. We can still use the term "electron" but now it refers to a wave function/state vector, with mass $m_e$, spin ½, and charge $–e$, not necessarily localized near a single point, rather than to a particle localized in a single small volume.

In addition, the assumption that there are localized particles can lead one astray in understanding the implications of experiments on entangled wave functions, whereas abandoning the concept of particles makes the explanation clear and simple. In the Bell-Aspect experiment, for example, one apparently has to assume instantaneous interactions at a distance if one supposes there are localized particles. But none of the puzzling properties seemingly implied by experiments on entangled wave functions arise if only the wave function exists; the correlations—sometimes long-range—built into the entangled wave functions lead directly to the experimental results.

Finally, the uncertainty principle becomes just a mathematical theorem about properties of the wave function rather than being a mysterious bound on the measurements of the momentum and position of "a particle."

Thus the unsolved mysteries that quantum mechanics gives rise to do *not* include wave-particle duality, the perception of only one version of reality, the results of entangled wave function experiments, the uncertainty principle, or localized effects from a spread-out wave function because these follow directly from the mathematics. Instead the primary mystery is the origin of the probability law [23,24]. In addition, since it is awkward to have the kets $|m, E, p, S, s_z, Q\rangle$ stand for the states of particles when there is no evidence for particles, a second major mystery is the nature of the kets. What do they represent [25]? If it is the states of matter, what is the nature of that matter?

## Appendix A. Comparison with the decoherence work of Zeh.

We will briefly compare our approach to the decoherence method of Zeh. In one sense, our use of orthogonality might be viewed as simply a different approach to the classicality problem. But there are also points where our treatment appears to offer an advantage.

1. The isolation of the different branches—dynamical independence—and thus the fact that each version of the observer perceives only what is on her branch, follows directly and easily in the orthogonality approach. The state of the environment and its coupling to the system are irrelevant in the orthogonality derivation (in those cases where different outcomes correspond to different perceptions by the versions of the observer), but the environment is critical in the decoherence approach.

2. Once dynamical independence is established, Zeh takes it as an *axiom* that only one version is "consciously" perceived; "So we may axiomatically identify these individual *component* states of the observer with states of consciousness (novel psycho-physical parallelism)." [26] (his italics). That is, he solves the preferred basis problem by assumption. In our treatment, on the other hand, we make the weak, eminently plausible assumption that the observer's perceptions must be communicable, and then *derive* that more than one classical version of reality cannot be perceived by any version of the observer.

3. Point 2—the perception of a single, classical version of reality—is critical for the derivation of (1) the perception of a single grain of film being exposed by a spread-out wave function, (2) the perception of particle-like trajectories, and (3) the perception of a localized object even though the object has a spread-out c.m. wave function. It is also critical for the justification of the rationale for superselection rules. Thus Zeh's derivations of these four points, based on his axiomatic assumption, are not as strong as our derivations, which are based on the weaker communicability assumption.

4. Finally, Zeh does not mention that the particle-like properties of mass, energy, momentum, spin and charge can logically, through group representation theory, be attributed to the state vectors (Sec. 3). These properties are an important aspect of the "classicality" of the physical world.

## Appendix B: No-Interaction Rule.

We show here that the orthogonality of the different detector states prevents interactions of any kind between the different branches. Thus no information on what happens on one branch can be transmitted to the version of the observer on a different branch.

To show this, suppose we do an experiment on an atomic system that has two possible outcomes. After the detector detects but before the observer observes, the state vector is (with 0 standing for no detection)

$$|\Psi(1)\rangle = \sum_{i=1}^{2} a(i)|i\rangle|D(i)\rangle|Obs(0)\rangle \qquad (B1)$$

The detector has a rotating dial that reads 1 on branch 1, and 2 on branch 2, so the atomic wave functions making up the detector are non-zero near one location on branch 1 and non-zero near a different location on branch 2, with the wave functions of the different versions of a particular atom being non-overlapping. Thus

$$\langle D(i)|D(j)\rangle = \delta_{ij} \tag{B2}$$

To see the impossibility of an interaction between branches, suppose version 1 of the detector sends out a photon and part of the photon wave function migrates in time to branch 2 so that (leaving out the atomic and observer states)

$$U(t)[|D(1)\rangle|\text{ph}\rangle] = a|D(1)\rangle|\text{ph}\rangle + b|D(2)\rangle|\text{ph}\rangle \tag{B3}$$

However, any reasonable, relevant U(t) does not materially change the positions of the atoms in the dial (that is, it does not change the setting of the dial). Thus if we take the scalar product of $\langle D(2)|$ with Eq. (B3), we get

$$\begin{aligned}\langle D(2)|U(t)[|D(1)\rangle|\text{ph}\rangle] &= 0 \\ &= b\langle D(2)|D(2)\rangle|\text{ph}\rangle\end{aligned} \tag{B4}$$

where the 0 came from the fact that U(t) leaves the atoms in location 1 so the scalar product with a bra having atoms in location 2 is zero. But since the scalar product on line 2 is non-zero, b must be 0. That is, the photons given off by detector 1 must stay on branch 1 (where the detector reads 1). It can therefore not interact with the versions of the detector and observer on branch 2.

We can do this more generally. We label the states by |branch i, det i⟩. Then U(t) |branch i, det i⟩ is still a state that has reading det i because U(t), by hypothesis, does not change the reading. That is, the time-evolved state has no admixture of a state with reading det j, so

$$\langle \text{branch j, det j}|U(t)|\text{branch i, det i}\rangle = 0,\ j \neq i \tag{B5}$$

Thus there can be no interaction between states having different readings on the detectors. This implies the i version of the observer can have no knowledge of the state of affairs on the j branch; it can only perceive what happens on the i branch.

The same reasoning can be used if the detector consists of grains of film, with the unexposed and exposed states orthogonal. In fact, the reasoning works whenever the different detector/recorder states are orthogonal, which should be true in all cases where there is a perceptual difference between the detector/recorder readings for the various outcomes.

## References and Notes.

distance), one could possibly account for the results. But there is no evidence this is the correct explanation.

[23] The probability mystery could conceivably be solved by a particle-like hidden variable theory. But there is *no evidence* that the existence of particles (as opposed, say, to collapse or something else) provides the solution to the probability mystery. We also note that if the probability mystery is solved by either particles or hidden variables or collapse, then perception of only one version of reality and localization both have two causes—the laws of pure quantum mechanics *and* one of the three modifications. Nature is not normally so extravagant.

[24] Zurek [3] claims to have derived the $|a_i|^2$ probability law from pure quantum mechanics by the use of auxiliary experiments. But he assumes the probability of "the observer" perceiving a particular outcome is the same whether or not the auxiliary experiments are done. That assumption may be permissible if one is arguing from a "physical" point of view (which is dangerously close to the view that there is a single, actual outcome instead of many simultaneous outcomes). But I do not believe it is a valid assumption when one's only basis for argument is the mathematics of pure quantum mechanics. In view of the mathematics—every version is perceived on every run, independent of the coefficients—it is at least as valid to conjecture that the probability of perception is 1/N, independent of the coefficients, where N is the number of observer states. His "auxiliary" assumption does not follow from the mathematics.

[25] See Blood, C.: Group representational clues to a theory underlying quantum mechanics, arXiv:quant-ph/0903.3160 (2009) for a conjecture on what the kets might represent.

[26] Zeh, H. D.,arXiv:quant-ph/9506020v3 (2002).